# ERROR-FREE MILESTONES IN ERROR PRONE MEASUREMENTS

By Dylan S. Small and Paul R. Rosenbaum[1]

*University of Pennsylvania*

A predictor variable or dose that is measured with substantial error may possess an error-free milestone, such that it is known with negligible error whether the value of the variable is to the left or right of the milestone. Such a milestone provides a basis for estimating a linear relationship between the true but unknown value of the error-free predictor and an outcome, because the milestone creates a strong and valid instrumental variable. The inferences are nonparametric and robust, and in the simplest cases, they are exact and distribution free. We also consider multiple milestones for a single predictor and milestones for several predictors whose partial slopes are estimated simultaneously. Examples are drawn from the Wisconsin Longitudinal Study, in which a BA degree acts as a milestone for sixteen years of education, and the binary indicator of military service acts as a milestone for years of service.

## 1. Introduction: strong, valid instrumental variables from error-free milestones.

1.1. *Error-free milestones.* A fallible measure contains an error-free milestone if there is some value of the measure, called the milestone, such that, despite errors of measurement, the measurement is always on the correct side of the milestone. Error-free milestones arise in a variety of ways. It may happen that a nonnegative quantity may contain errors when it is strictly positive, but a zero is truly and exactly a zero; an example involving duration of exposure to anesthetics is discussed in Section 1.2. If a scale is defined in terms of many item responses, then for some possible definitions of the scale, an error free item yields an error free milestone; an example involving a scale of exposure to combat in Vietnam is discussed in Section 1.3.

Received May 2008; revised October 2008.

[1]Supported in part by a grant from the Methodology, Measurement and Statistics Program of the U.S. National Science Foundation.

If deceit is distinguished from error, then the concept of an error-free milestone (as distinct from a deception free milestone) is relevant to responses to questions; see Section 1.4.

In practice, an error-free milestone is a model intended to approximate situations in which errors that respect the milestone are commonplace and errors that cross the milestone are extremely infrequent. In Section 1.2 imprecision in recording the duration of anesthesia respects a milestone at zero, whereas failing to bill for an operation causes an error that crosses the milestone; however, there are strong disincentives for the latter error.

1.2. *Inhaled anesthetics and neurodegenerative disorders.* Measurable cognitive dysfunction may occur in perhaps 20% of patients one week after surgery with an inhaled anesthetic [Johnson et al. (2002)], but long term effects in humans have not, so far, been demonstrated. Eckenhoff et al. (2004) provide in vitro laboratory evidence suggesting that the anesthetics halothane and isoflurane enhanced cellular changes associated with the development of neurodegenerative disorders such as Alzheimer and Parkinson disease.

Mounting a large scale, long term study in humans faces several significant obstacles, including (i) measurement of the duration of anesthetic exposure, (ii) measurement of neurodegenerative outcomes, and (iii) confounding of the need for surgery with effects of anesthetics given during surgery. Jeffrey Silber, Roderic Eckenhoff and one of the authors (Rosenbaum) have proposed to use data from Medicare as the basis for such a study. Medicare is the program of the U.S. government which provides publicly financed health care to people of age 65 or greater. With some exceptions, doctors and hospitals bill Medicare for services provided to the elderly, and these Medicare claims create a national record of health care for Medicare recipients.

If you fall in a certain way and break your hip, it is likely that you will need hip surgery requiring prolonged anesthesia; for Medicare recipients, these events will be recorded in Medicare claims. If you fall in a slightly different way and break your pelvis, it is likely that your condition will be treated without surgery, and hence without inhaled anesthesia; these events, also, will be recorded in Medicare claims. Comparing patients who broke either a hip or a pelvis is an example of using differential treatment effects to remove confounding from generic biases [Rosenbaum (2006)].

Like lawyers, and unlike surgeons, anesthesiologists bill for their time. Anesthesiologists submit a bill to Medicare which records the duration of anesthetic care. Silber et al. (2007) compared the times recorded in these bills to times obtained by chart abstraction for 1931 patients in Pennsylvania. The bills were typically in close agreement with the chart abstractions, with a median absolute difference of five minutes, but as seen in the quantile–quantile plot in Figure 2 of Silber et al. (2007), the distribution is approximately symmetric with extremely long tails, with more than 1% of bills discrepant by more than an hour. The cause of these large discrepancies is not known, and could conceivably be errors in abstraction rather than in bills; however, we suspect that our algorithm for record linkage sometimes makes a few gross errors, possibly due to errors in dates on bills. Silber et al. (2007) conclude that anesthesia bills can be used to gauge anesthesia duration, providing robust methods are used to prevent the long tails from having inappropriate influence.

Although the anesthesia bills measure anesthesia duration with moderate but long tailed error, it is virtually certain that a patient who did not have surgery had no exposure to inhaled anesthetics. In other words, the nonzero anesthesia duration for a broken hip will contain error, but the zero duration for a broken pelvis will truly be zero, creating an error-free milestone at zero.

1.3. *Scales in which certain levels may be verified using administrative records.* Lund et al. (1984) created a seven point scale of the degree of exposure to combat during the Vietnam War. Certain points on the scale can be determined objectively from military records; others depend on self-report. For instance, "in military service during 1965–1975" by itself scores a 0, whereas that combined with "stationed in Vietnam" scores a 1, while both of these together with "saw injury or death of U.S. Serviceman" scores a 2, and so on, and "wounded in combat" scores a 5. Military records indicate when and where an individual has served and whether the individual was wounded in combat, but there is no record of whether an individual saw the injury or death of a U.S. Serviceman. A misstatement by an individual may result in erroneous placement on the scale, but only within the milestones created by the scale's dependence, at certain points, on objective records.

Expressed more abstractly, it is common to combine several oriented pieces of information or items to form a scale. Here the scale is the degree of exposure to combat and the items are such events as "wounded in combat." With $m$ binary items, the $2^m$ possible patterns of item responses are partially ordered, for instance, a person who is positive for items 1 and 2 and for no other items is at least as high in the partial order as a person who is positive for item 2 and no other item, etc. In rare instances, the patterns that actually occur form a linear order or Guttman scale, so only $m+1$ patterns of the $2^m$ possibilities actually occur. More commonly, the definition of the scale imposes a linear order that is compatible with (i.e., is a linear extension of) the partial order on the $2^m$ possible patterns. If some of the items are error free and others are error prone, then it is always possible to define the linear order or scale so that it gives lexicographic priority to at least one of the error-free items, thereby creating an error-free milestone; see the discussion of the lexicographic sum of partial orders in Trotter (1992), page 24. Whether or not such a scale will be reasonable as a scale obviously depends upon the content of the specific items involved, but the mere existence of scales with error-free milestones is guaranteed by the existence of at least one error-free item.

As noted by Dee, Evans and Murray (1999), in longitudinal data for research in education and labor economics, it is increasingly common to combine transcripts from educational institutions with survey questionnaires. Although this does not appear to have been done as yet, in parallel with Lund et al. (1984), one could create educational scales anchored by milestones determined from transcripts, for instance, receipt of particular academic degrees.

1.4. *Milestones to anchor memory.* Measurements that describe people are often obtained by asking them questions. How many years of education do you have? How long did you serve in the U.S. military? How many cigarettes do you smoke per day? To what extent are you prone to violent behavior? In asking such questions, an investigator hopes that the respondent can remember the answer, and can express the answer in a manner consistent with the investigator's operational definitions, but these hopes are not always realized. Aspects of questionnaire design are discussed by Sudman and Bradburn (1986), Lyberg (1997) and Tourangeau, Rips and Rasinski (2000).

For instance, by "years of education," most investigators mean "grades successfully completed," not years spent trying. Imagine a person who dropped out of high school in the middle of tenth grade, having repeated grades three and seven. Such a person might think of this as twelve years of education (ten plus two), whereas the investigator might intend this to be classified as successful completion of grades one through nine, or nine years of education. Similarly, a person who achieves a BA degree with three years of college, a summer session after the freshman year and some advanced placement credit might report fifteen years of education, whereas the investigator might intend to credit sixteen years of education for achievement of a BA.

A respondent may intend to report accurately, but may fail to do so because of lapses of memory and uncertainties about the intended meaning of the question. Certain events, however, are easy to remember and unambiguous in question and answer: they are events punctuated by public ceremony, official sanction, public documents, and by *kinds* of behavior rather than *degrees* of behavior. An honest, sober, mentally competent respondent is unlikely to err in response to the following questions: Do you have a high school degree or high school equivalency degree? Did you ever serve in the U.S. military? Have you smoked any part of at least one cigarette in the last seven days? Have you ever been convicted for assault? If scales of behavior are defined in terms of such unambiguous milestones, and if questioning is organized to ensure that the milestone is respected in responses, then the milestones may be measured with negligible error, despite continued errors at points in the scale between milestones.

1.5. *Outline.* Our purpose here is formalize these considerations, showing how error-free milestones permit estimation of slopes for the true but unknown error-free measurements. Speaking informally, almost by definition of the scale itself, an error-free milestone creates a strong and valid instrumental variable, so that location with respect to the milestone is related to the true measurement but is uncontaminated by measurement error; see Section 2 for formal definitions and results. The inferences are nonparametric and robust, and in the simplest cases they are exact and distribution free. In Section 2 the most common and simplest case is discussed, namely, a single milestone for a single variable, first for matched pairs using Wilcoxon's signed rank test, then for matched sets formed by full matching using a generalization of the signed rank test. In Section 4, multiple milestones are considered, including several milestones for one predictor, single milestones for each of several predictors, or several milestones for each of several predictors. The theory in Sections 2 and 4 is applied in Sections 3 and 5, respectively, to the example in Section 1.6.

Our use of the term instrumental variable departs slightly from the traditional definition, which is stated in terms of covariances; see Cheng and Van Ness (1999), Section 4, for review of the traditional definition. The literature on correcting for measurement error is extensive; see also Kendall and Stuart (1973), Section 29, Fuller (1987), Brenner and Gefeller (1993) and Carrol, Ruppert and Stefanski (1995) for several perspectives. The method discussed in Rosenbaum (2005) may be viewed as a special case in which the error-free milestone occurs between dose zero and all positive doses, in which case it was possible to correct for measurement error using controls known to have received dose zero of a treatment. The notion of error-free milestones is substantially more general, however, in that all doses may be affected by errors, and several milestones may be available.

1.6. *Years of education in the Wisconsin Longitudinal Study.* Traditional questions in sociology and labor economics concern the effects of additional schooling or of service in the military. The Wisconsin Longitudinal Study (WLS) provides especially detailed information, including an IQ test score from high school, and several measures of education. For two of the many empirical studies based on the WLS, see Singer et al. (1998) and Warren, Sheridan and Hauser (2002). We focus on the 3738 men with wages of at least $100 in 1974. The WLS began its data collection with surveys in the senior year of high school, which in the U.S. is conventionally recorded as 12 years of education, with kindergarten and preschool ignored. In WLS, the variable edyrcm is self-reported years of education beyond high school, which

we use in the form SR = edyrcm + 12, where SR signifies "self-report." The second measure, edeqyr, is a scaled measure of education based on equivalent degrees actually earned (DS for "degree scaled"), for example, 16 years for a BA, 20 years for a Ph.D., etc. Using DS, we create a binary indicator of whether the individual reports having a BA degree. These two measures of education, SR and DS, often differ by a few years, but they are in substantial agreement at 16 years of education for the BA. Although not collected in precisely the manner suggested in Section 1.4, to a close approximation, SR does seem to have the BA degree as an error-free milestone: in SR, all but $29/3738 = 0.008 < 1\%$ of the men reported less than 16 years of education if no BA was received or at least 16 years of education if a BA was received.

Figure 1 contrasts three measures of education in the WLS, including the degree scaled education, DS, and the self reported education, SR. The DS and SR differ for $470 = 12.6\%$ of the men, mostly by one year, but discrepancies as large as seven years do occur. We assumed that the report of a BA or not was accurate, and created a third measure, the adjusted self report or SRa, which minimally altered the $29/3738 = 0.008 < 1\%$ of the men whose self-reported years were inconsistent with 16 years for the BA. Specifically, two men who reported a BA with 15 years of education were credited with 16 years of education, 24 men who reported no BA with 16 years of education and 3 men who reported no BA with 17 years of education were credited with 15.999 years of education. This adjustment would not be necessary if the questionnaire forced compliance with the milestone. Of course, because only small changes were made to 29/3738 records, in Figure 1, SR and SRa are indistinguishable, but both differ somewhat from DS. For the purpose of illustration in the current paper, we act as if SRa were a fallible measure of DS with a milestone at 16 years.

In Sections 3 and 5, we estimate the relationship between log earnings in 1974 and education correcting for errors of measurement in self-reported education using the BA degree as a milestone for 16 years of education.

## 2. Inference using one milestone.

2.1. *Doses measured with random errors.* There are $I$ matched sets or strata, $i = 1, \ldots, I$, matched exactly on covariates $\mathbf{x}$, and matched set $i$ contains $n_i \geq 2$ individuals, $j = 1, \ldots, n_i$. The $j$th individual in set $i$ has covariates $\mathbf{x}_{ij}$, true but unobserved dose $d_{ij}$, fallible observed dose $D_{ij}$, and outcome $Y_{ij}$. Here, $\mathbf{x}_{ij}$ and $d_{ij}$ are viewed as fixed, perhaps fixed by conditioning as in a regression model, but $D_{ij}$ and $Y_{ij}$ are random variables, in part because $D_{ij}$ measures $d_{ij}$ with an error of measurement. Because the matching is exact, $\mathbf{x}_{ij} = \mathbf{x}_{ij'}$ for all $i$, $j$, $j'$. See Cochran (1968) for some discussion of the consequences of close but inexact control for $\mathbf{x}$.

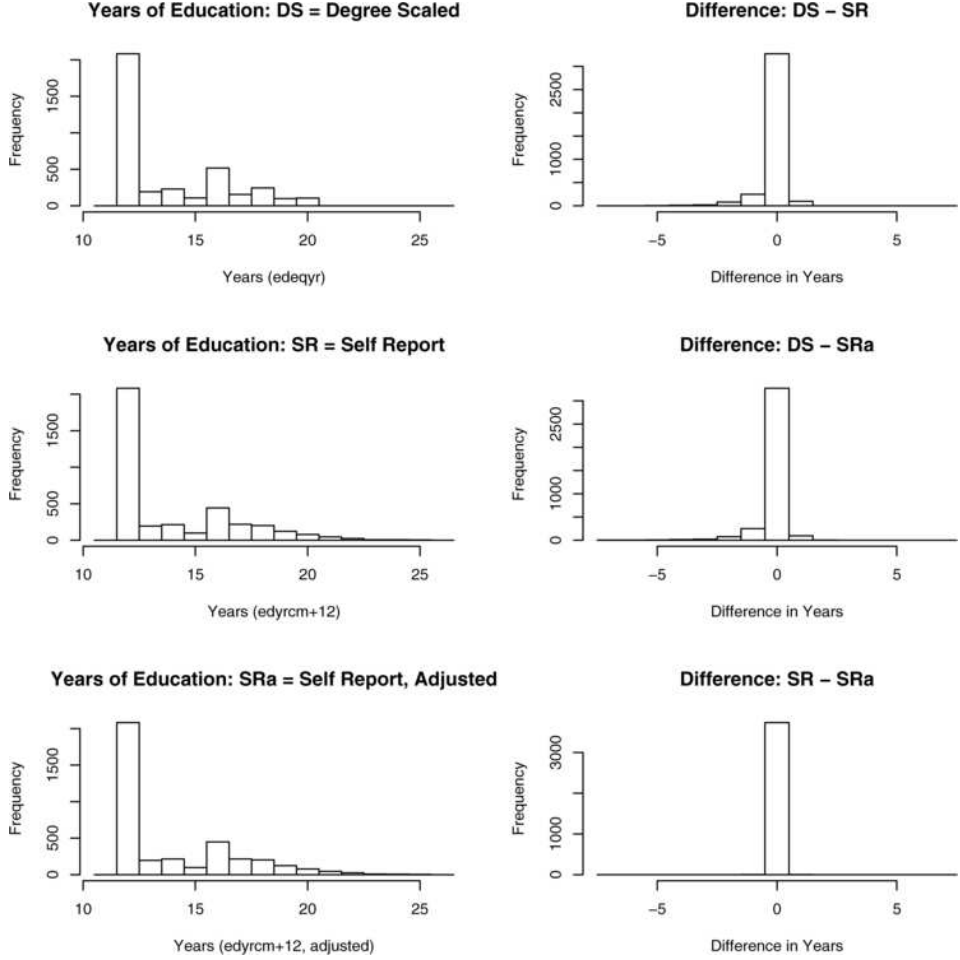

FIG. 1. *Three measures of education compared: degree scaled (DS), self-report (SR) and self-report minimally edited to be compatible with a BA = 16 years (SRa). There are a fair number of small discrepancies between DS and SR, and a very small number of larger discrepancies (up to seven years). There are only 29/3738 discrepancies between SR and SRa, of which 27 are about one year, and two are two years.*

Let $\mathcal{C}$ be the set of continuous distribution functions on the real line, and let $\mathcal{S}$ be the subset of continuous distribution functions on the line that are symmetric about zero. The true but unknown dose $d_{ij}$ is assumed to be linearly related to $Y_{ij}$,

$$Y_{ij} = \lambda(\mathbf{x}_{ij}) + \beta d_{ij} + \varepsilon_{ij}, \qquad \varepsilon_{ij} \stackrel{\text{i.i.d.}}{\sim} G \in \mathcal{C}, \tag{1}$$

where $\beta$ is the parameter to be estimated, and $\lambda(\cdot)$ is an unknown function. The fallible, observable dose $D_{ij}$ measures the true but unknown dose $d_{ij}$ with errors $\xi_{ij}$ that are symmetric about zero, are mutually independent, and independent of the $\varepsilon_{ij}$,

$$(2) \qquad D_{ij} = d_{ij} + \xi_{ij}, \qquad \xi_{ij} \sim F_{d_{ij}} \in \mathcal{S},$$

so the distribution of measurement errors, $F_{d_{ij}}$, varies with the true dose $d_{ij}$, but $D_{ij}$ is always symmetrically distributed about its center or median, namely, $d_{ij}$. So far, (1) and (2) slightly generalize the traditional errors-in-variables regression model [Wald (1940), Neyman and Scott (1951), Madansky (1959), Kendall and Stuart (1973), Section 29, Fuller (1987), Section 1, Cheng and Van Ness (1999)], notably because the distribution $F_{d_{ij}}$ of errors $\xi_{ij}$ need not be the same for all true doses $d_{ij}$. To say that $D_{ij}$ measures $d_{ij}$ with error, there must be some sense in which the error $D_{ij} - d_{ij} = \xi_{ij}$ is typically zero, and the symmetry of the distribution $F_{d_{ij}}$ of $\xi_{ij}$ about zero in (2) is one such sense. Later, we remove the assumption of symmetry in (2), replacing it by the assumption that $E(\xi_{ij}) = 0$, but for now $\xi_{ij}$ is symmetric about zero. For instance, $\xi_{ij}$ might have a rescaled and relocated symmetric beta distribution with median zero and with a range and a shape that might vary in some way with $d_{ij}$. If $D_{ij} - d_{ij}$ is centered, say, at positive value, then the doses are systematically biased, and the methods we propose apply to random errors of measurement but not to systematic biases.

It is well known that $\beta$ is not identified under models (1) and (2). Indeed, even if one assumed much more, say, that $\lambda(\mathbf{x}_{ij}) = \alpha$ for all $\mathbf{x}_{ij}$, and $\varepsilon_{ij} \sim N(0, \sigma_\varepsilon^2)$, $\xi_{ij} \sim N(0, \sigma_\xi^2)$, $d_{ij} \sim N(\mu_d, \sigma_d^2)$ with unknown $\sigma_\varepsilon^2 > 0$, $\sigma_\xi^2 > 0$, $\sigma_d^2 > 0$ and $\alpha$, then: (i) there would be no consistent estimate of $\beta$; (ii) the likelihood function would have a ridge rather than a unique maximum; (iii) least squares regression of $Y_{ij}$ on $D_{ij}$ would be consistent for $\beta \sigma_d^2 / (\sigma_d^2 + \sigma_\xi^2) \neq \beta$, where $D_{ij}$ has reliability $\sigma_d^2 / (\sigma_d^2 + \sigma_\xi^2) < 1$ as a measure of $d_{ij}$; see Cheng and Van Ness (1999), Section 1.2.1.

2.2. *Definition of a milestone.* The number $\kappa$ is defined to be an error-free milestone, or briefly a milestone, for $(D_{ij}, d_{ij})$ if

$$(3) \quad D_{ij} < \kappa \iff d_{ij} < \kappa, \qquad D_{ij} \geq \kappa \iff d_{ij} \geq \kappa, \qquad \forall i, j.$$

In the WLS example in Section 1.6, with $\kappa = 16$ years of education, (3) says that a respondent might misreport $d_{ij}$ years of education as $D_{ij}$ years of education because of a lapse of memory or a miscommunication about the investigator's operational definition of what counts as a year of education, but an honest, mentally competent respondent could not misunderstand or forget the answer to the question: "Did you receive a BA degree?"

Obviously, a milestone at $\kappa$ in (3) places a restriction on the range of the distribution $F_{d_{ij}}$ of the error of measurement $\xi_{ij}$. If (3) is true with $\kappa = 16$ in Section 1.6, then a man who reports $D_{ij} = 18$ years of education has at least $d_{ij} \geq \kappa = 16$ years of education, so he exaggerates his education by at most two years, $\xi_{ij} = D_{ij} - d_{ij} \leq 2$. Similarly, a man who reports $D_{ij} = 14$ years of education has at most $d_{ij} < \kappa = 16$ years of education, so he understates his education by at most $\xi_{ij} = D_{ij} - d_{ij} > -2$ years. In general, if $D_{ij} \geq \kappa$, then $d_{ij} \geq \kappa$ so that $\xi_{ij} = D_{ij} - d_{ij} \leq D_{ij} - \kappa$, whereas if $D_{ij} < \kappa$, then $d_{ij} < \kappa$ so that $\xi_{ij} = D_{ij} - d_{ij} > D_{ij} - \kappa$. This range restriction is respected by various parametric families of distributions $F_{d_{ij}}$ for $\xi_{ij}$ in (2) which are symmetric about zero, $F_{d_{ij}} \in \mathcal{S}$, including the symmetric beta distributions relocated and rescaled to have median zero with support contained in the interval $[-|d_{ij} - \kappa|, |d_{ij} - \kappa|]$.

2.3. *A basic property.* Consider testing the hypothesis $H_0 : \beta = \beta_0$ in (1) and (2) using the error-free milestone (3). Recall that the matching on $\mathbf{x}$ is exact, $\mathbf{x}_{ij} = \mathbf{x}_{ik}$. If a matched set $i$ contains an individual $j$ with $D_{ij} \geq \kappa$ and another individual $k$ with $D_{ik} < \kappa$, then compute

$$
\begin{aligned}
Q_{ijk}^{(\beta_0)} &= (Y_{ij} - \beta_0 D_{ij}) - (Y_{ik} - \beta_0 D_{ik}) \\
&= \beta(d_{ij} - d_{ik}) - \beta_0(D_{ij} - D_{ik}) + (\varepsilon_{ij} - \varepsilon_{ik}) \\
&= (\beta - \beta_0)(d_{ij} - d_{ik}) - \beta_0(\xi_{ij} - \xi_{ik}) + (\varepsilon_{ij} - \varepsilon_{ik}).
\end{aligned}
\tag{4}
$$

Because $\kappa$ is a milestone in (3), $d_{ij} - d_{ik} > 0$ in (4). Also, because the $\xi_{ij}$, $\xi_{ik}$, $\varepsilon_{ij}$, $\varepsilon_{ik}$ are mutually independent with distributions satisfying the conditions in (1) and (2), the quantity $-\beta_0(\xi_{ij} - \xi_{ik}) + (\varepsilon_{ij} - \varepsilon_{ik})$ in (4) has a continuous distribution symmetric about zero. If $H_0 : \beta = \beta_0$ were true, then $Q_{ijk}^{(\beta_0)}$ in (4) would be symmetrically distributed about zero. If $H_0 : \beta = \beta_0$ were false with $\beta > \beta_0$, then $Q_{ijk}^{(\beta_0)}$ would be symmetrically distributed about a positive quantity, whereas if $\beta < \beta_0$, then $Q_{ijk}^{(\beta_0)}$ would be symmetric about a negative quantity.

The symmetry of $Q_{ijk}^{(\beta_0)}$ about $(\beta - \beta_0)(d_{ij} - d_{ik})$ also holds under certain variations of the model in (1) and (2). In particular, the $\varepsilon_{ij}$ need not all have the same distribution $G \in \mathcal{C}$; rather, they could have different distributions that are symmetric about zero, $\varepsilon_{ij} \sim G_{d_{ij}} \in \mathcal{S}$, and then $Q_{ijk}^{(\beta_0)}$ would still be symmetric about $(\beta - \beta_0)(d_{ij} - d_{ik})$. In Section 1.6, for instance, a person with $d_{ij} = 18$ years of education might have either a law degree or a masters degree in art history, so the $\varepsilon_{ij}$ for wages $Y_{ij}$ might be more variable at $d_{ij} = 18$ than at $d_{ij} = 12$, so $G_{18}$ might be more dispersed than $G_{12}$, but providing $\varepsilon_{ij} \sim G_{d_{ij}} \in \mathcal{S}$, in (4) the quantity $Q_{ijk}^{(\beta_0)}$ is symmetric about $(\beta - \beta_0)(d_{ij} - d_{ik})$.

If one replaces all assumptions of symmetry of $\xi_{ij}$ or $\varepsilon_{ij}$ by the assumption that $E(\xi_{ij}) = E(\varepsilon_{ij}) = 0$, then $Q_{ijk}^{(\beta_0)}$ in (4) has expectation $E\{Q_{ijk}^{(\beta_0)}\} = (\beta - \beta_0)(d_{ij} - d_{ik})$, and, in particular, $E\{Q_{ijk}^{(\beta_0)}\} = 0$ if $H_0 : \beta = \beta_0$ is true.

2.4. *Inference with matched pairs.* Suppose that $\kappa$ is a milestone (3) for $(D_{ij}, d_{ij})$ and $I$ pairs, $n_i = 2$, $i = 1, \ldots, I$, are matched exactly for $\mathbf{x}_{ij}$ with the additional requirement that $D_{i1} \geq \kappa > D_{i2}$, or, equivalently, the requirement that $d_{i1} \geq \kappa > d_{i2}$. In Section 1.6 this would mean pairing someone with at least a BA to someone with less than a BA. Although $(D_{i1}, D_{i2}) = (d_{i1} + \xi_{i1}, d_{i2} + \xi_{i2})$ is a random quantity because $(\xi_{i1}, \xi_{i2})$ is random, the event $D_{i1} \geq \kappa > D_{i2}$ is determined by $(d_{i1}, d_{i2})$, which is fixed. In Section 1.6 this would mean that, although there are random errors in reported years of education $(D_{i1}, D_{i2})$, the pairing of someone with at least a BA to someone with less than a BA is made without error: in each pair, the person claiming to have a BA has one, and the person claiming not to have a BA does not have one.

To test $H_0: \beta = \beta_0$ in (1) and (2), calculate the $I$ mutually independent differences, $Q_{i12}^{(\beta_0)} \doteq (Y_{i1} - \beta_0 D_{i1}) - (Y_{i2} - \beta_0 D_{i2})$. The $Q_{i12}^{(\beta_0)}$ are symmetrically distributed about $(\beta - \beta_0)(d_{i1} - d_{i2})$ by (4), where $d_{i1} - d_{i2} > 0$ because the pairing ensured $d_{i1} \geq \kappa > d_{i2}$. Let $T_{\beta_0}$ be Wilcoxon's signed rank statistic [e.g., Hettmansperger and McKean (1998), Section 1] computed from $Q_{i12}^{(\beta_0)}$; that is, rank the $|Q_{i12}^{(\beta_0)}|$ from 1 to $I$, and let $T_{\beta_0}$ be the sum of the ranks for which $Q_{i12}^{(\beta_0)} > 0$. If $H_0: \beta = \beta_0$ is true, then $\text{sign}\{Q_{i12}^{(\beta_0)}\}$ and $|Q_{i12}^{(\beta_0)}|$ are independent, where $\text{sign}(a) = 1$, 0, or $-1$ as $a > 0$, $a = 0$, $a < 0$; see Wolfe (1974), Corollary 2.1. So if $H_0: \beta = \beta_0$ is true, then the conditional distribution of $T_{\beta_0}$ given the $|Q_{i12}^{(\beta_0)}|$ is the usual exact distribution of Wilcoxon's signed rank statistic, namely, the distribution of the sum of $I$ independent random variables taking values $i$ or 0 each with probability $\frac{1}{2}$, $i = 1, \ldots, I$. Therefore, $T_{\beta_0}$ yields an exact, distribution free test of $H_0: \beta = \beta_0$. If $\beta > \beta_0$, the $Q_{i12}^{(\beta_0)}$ are symmetric about $(\beta - \beta_0)(d_{i1} - d_{i2}) > 0$, so the test based on $T_{\beta_0}$ is consistent against $H_1: \beta > \beta_0$ under mild conditions on the limiting behavior of the fixed $d_{ij}$'s and of the $F_{d_{ij}}$ as $I \to \infty$. Similarly, the test is consistent against $H_1: \beta < \beta_0$. A $1 - \alpha$ confidence set for $\beta$ is formed by inverting the test, that is, as the set of hypotheses $H_0: \beta = \beta_0$ not rejected by a level $\alpha$ test. Because $D_{i1} \geq \kappa > D_{i2}$, the difference $Q_{i12}^{(\beta_0)}$ is strictly decreasing as a function of $\beta_0$, so the signed rank statistic $T_{\beta_0}$ is monotone decreasing as a function of $\beta_0$, which implies that this confidence set is an interval. Under $H_0: \beta = \beta_0$, the null expectation of the signed rank statistic is $I(I+1)/4$. The Hodges–Lehmann (1963) point estimate $\widehat{\beta}$ of $\beta$ is the "solution" to the estimating equation, $T_{\widehat{\beta}} = I(I+1)/4$, in a sense that will now be described. Because the rank statistic $T_{\beta_0}$ takes many, small discrete steps downward as $\beta_0$ increases continuously, there is either a unique value, $\widehat{\beta}$, of $\beta_0$ where $T_{\beta_0}$ passes $I(I+1)/4$, or there is an interval of values of $\beta_0$ where $T_{\beta_0} = I(I+1)/4$, in which case the "solution" $\widehat{\beta}$ is defined to be the midpoint of this interval.

An alternative estimator uses sample means rather than rank statistics. Write $\overline{Q}_{12}^{(\beta_0)} = (1/I) \sum Q_{i12}^{(\beta_0)} = (\overline{Y}_1 - \beta_0 \overline{D}_1) - (\overline{Y}_2 - \beta_0 \overline{D}_2)$, where $\overline{Y}_1 = (1/I) \sum Y_{i1}$, etc. Assume in this paragraph only that the i.i.d. $\varepsilon_{ij}$'s have finite variance and that the $\xi_{ij}$'s, which are not i.i.d., have uniformly bounded variances. If $H_0: \beta = \beta_0$ is true in (1) and (2), then $E\{\overline{Q}_{12}^{(\beta_0)}\} = 0$. The estimating equation $\overline{Q}_{12}^{(\widetilde{\beta})} = 0$ has solution $\widetilde{\beta} = (\overline{Y}_1 - \overline{Y}_2)/(\overline{D}_1 - \overline{D}_2)$, which is Wald's (1940) estimator, or two-stage least squares, in a context that avoids the concerns raised by Neyman and Scott (1951). In $\widetilde{\beta}$, the denominator has positive expectation, $E(\overline{D}_1 - \overline{D}_2) > 0$ because $D_{i1} \geq \kappa > D_{i2}$, and $\widetilde{\beta}$ is consistent for $\beta$ under mild conditions on the limiting behavior of the fixed $d_{ij}$'s and of the $F_{d_{ij}}$ as $I \to \infty$. In parallel with the procedures above using the signed rank test, a one-sample $t$-statistic may be computed from the $Q_{i12}^{(\beta_0)}$. This $t$-statistic does not have a $t$-distribution, in part because the $\xi_{ij}$'s in $Q_{i12}^{(\beta_0)}$ are not i.i.d. Normal random variables, and their variances may change with $d_{ij}$. With i.i.d. Normal matched pair differences, the Pitman asymptotic relative efficiency of the signed rank statistic and the $t$-statistic is $3/\pi = 0.955$, and Sen's (1968) Theorem 2.2, result shows that the relative efficiency is always greater than or equal to $3/\pi$, often much greater than 1, with Normal distributions having unequal variances. In short, in this context, the signed rank statistic is robust to outliers, has a known finite sample null distribution, and has the possibility of superior efficiency relative to the $t$-statistic. The procedures based on means do have one advantage: unlike the signed rank statistic, they yield consistent inferences as $I \to \infty$, assuming $E(\xi_{ij}) = 0$ without the assumption that the $\xi_{ij}$ are symmetrically distributed about zero.

2.5. *Inference with matched sets.* In a full matching, each matched set with $n_i \geq 2$ individuals contains either 1 individual with $D_{ij} \geq \kappa$ and $n_i - 1$ individuals with $D_{ij} < \kappa$ or else $n_i - 1$ individuals with $D_{ij} \geq \kappa$ and 1 individual with $D_{ij} < \kappa$. Matched pairs, as in Section 2.4, and matching with a fixed number of controls are special cases of full matching. It can be shown that the stratification or matching that minimizes the total distance on **x** within matched sets is always a full matching, and an optimal full matching— one that minimizes the total distance within matched sets—may be constructed by solving a standard combinatorial optimization problem, technically known as minimum cost flow in a network [Rosenbaum (1991), Gu and Rosenbaum (1993), Hansen (2004, 2007), Hansen and Klopfer (2006)]. Such a matched set creates $n_i - 1$ differences $Q_{ijk}^{(\beta_0)}$ of the form (4); however, these $n_i - 1$ differences are now dependent because one of the $D_{ij}$'s, say, $D_{i1}$, appears in all $n_i - 1$ differences. If $H_0: \beta = \beta_0$ were true in (1) and ([2](#)), each $Q_{ijk}^{(\beta_0)}$ in ([4](#)) would be symmetric about zero, and the $n_i - 1$ differences would have a joint distribution with a form of reflection symmetry about **0** described by Sen and Puri ([1967](#)); specifically, $(Q_{i12}^{(\beta_0)}, \ldots, Q_{i,1,n_i}^{(\beta_0)})$ would have the same distribution as $(-Q_{i12}^{(\beta_0)}, \ldots, -Q_{i,1,n_i}^{(\beta_0)})$. If $H_0 : \beta = \beta_0$ were true, for any statistic that is a function of the $Q_{ijk}^{(\beta_0)}$, the reflection symmetry yields a null permutation distribution formed by changing the signs of the $I$ vectors $(Q_{i12}^{(\beta_0)}, \ldots, Q_{i,1,n_i}^{(\beta_0)})$ in all $2^I$ possible ways; see Sen and Puri ([1967](#)) and Rosenbaum ([2005](#)) for details. For instance, in Section [3.2](#) the usual Wilcoxon signed rank statistic is compared with this unusual permutation distribution which correctly allows for dependence in matched sets with $n_i > 2$; see Rosenbaum ([2005](#)) for a computational illustration.

## 3. An example with a single milestone: education and earnings.

3.1. *Full matching to control for IQ, parent's education and home town.* The method of Section [2](#) will be applied to the example in Section [1.6](#), using the BA degree as a milestone for 16 years of education in the self-reported years of education, SRa. We contrast the results with least squares and Huber's ([1981](#)), Section 7, $m$-estimation using SRa, ignoring measurement error and using a linear model for the covariates $\mathbf{x}_{ij}$. In $m$-estimation, we used the defaults for rlm in the MASS package in R. Under the simplest models for errors of measurement, we expect the slope estimates from least squares and $m$-estimation to be attenuated, or biased toward zero, and the estimate using the milestone to be consistent. The degree scaled measure of education, DS, used as the standard for comparison, is analyzed in a parallel manner. In practice, the simple measurement error models may be incorrect, and the methods differ in several ways, but the comparison serves as an illustration. In Section [3.1](#) the matching is described, while in Section [3.2](#) the estimated economic returns to additional education are compared.

In the Wisconsin Longitudinal Study in Section [1.6](#), there were 1124 men with a BA degree, and 2614 men without one, $3738 = 1124 + 2614$. The 1124 men with a BA were matched to 1124 men without a BA. The matching controlled for a four dimensional $\mathbf{x}$, whose coordinates were IQ in high school (specifically gwiiq_bm), father's education in years (edfa57q), mothers education in years (edmo57q) and the population size of the town in which the individual attended high school (pop15). Parental education was missing in whole or in part for 432 men, and an effort was made to match men with missing parental education to other men with missing parental education.

Pair matching is not feasible in these data, because the distributions of $\mathbf{x}$ are quite different for males with a BA and males without a BA. This is seen for IQ in Figure [2](#) which depicts the IQ's for the 1124 males with a BA and the 1124 highest IQ's for males without a BA among the 2614 males without a BA. Even the 1124 highest IQ's without a BA are too low to form an acceptable match. Moreover, these 1124 highest IQ's would constitute a poor match, in part because they ignore the other three covariates, and in part because some lower IQ's are needed to match to males with BA's having lower IQ's.

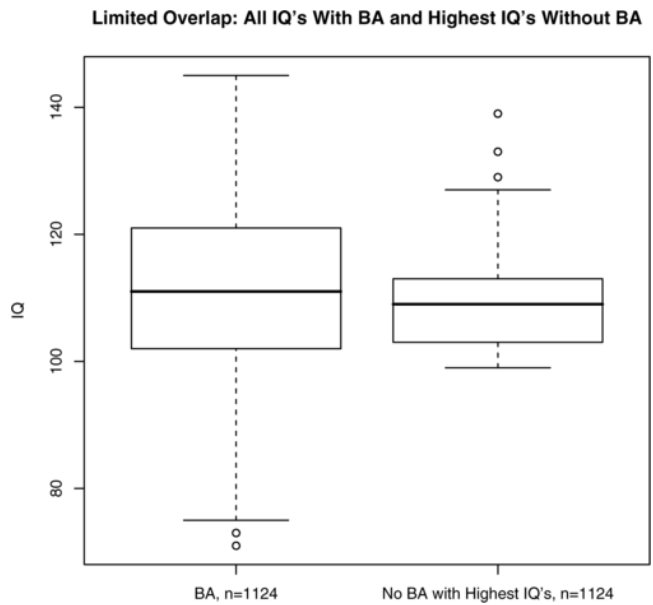

FIG. 2. *IQ scores for all 1124 males with a BA and for the 1124 highest IQ scores among the 2614 males without a BA. The figure shows that pair matching is not feasible, even if matching on IQ were the only objective.*

In place of pair matching, a full matching was performed, with a maximum 2-to-1 ratio, using all 1124 males with a BA and 1124 males without a BA. This means that a matched set might be a matched pair or a matched triple. A pair consists of a male with a BA and a male without a BA, and there were 239 such pairs. A triple may consist of either a male with a BA and two without a BA, or two with a BA and one without, and there were 295 triples of each type. That is, there were $829 = 239 + 295 + 295$ matched sets, containing $1124 = 239 + 295 + 2 \times 295$ males with a BA, and the same number without a BA. At higher IQ's, two men with a BA might be matched to one without a BA, with the reverse pattern at lower IQ's. As noted in Section 2.5, full matching is the form that minimizes distances within matched sets [Rosenbaum (1991)], and an implementation of optimal full matching is available in the optmatch package in R [Hansen (2004, 2007), Hansen and Klopfer (2006)]. Haviland, Nagin and Rosenbaum (2007), Appendix, present a general result about efficiency from matched sets with varied match ratios, and the 1–2 limit on imbalance is quite efficient. The distance used was the

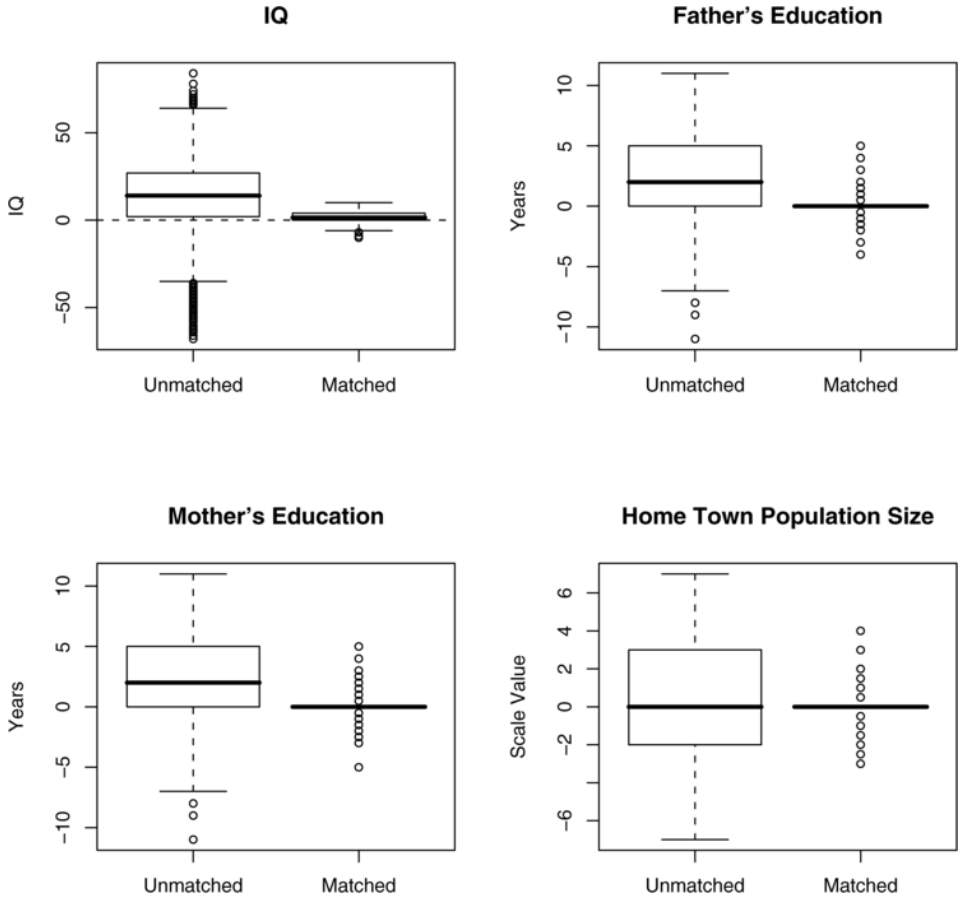

Fig. 3. *Four covariates before and after full matching. "Unmatched" refers to all pairwise differences, BA-minus-no-BA. "Matched" refers to the differences in means within 829 matched pairs or triples.*

Mahalanobis distance on the ranks of the four variables, with two additional variables containing binary indicators of missing parental education. Because the Mahalanobis distance is affinely invariant and missing indicators are included, any value may be substituted for missing values without altering the Mahalanobis distance. Figure 3 shows the four covariates before and after full matching. Each covariate is represented by a pair of boxplots, one before matching, the other after matching. The boxplot before matching compares the 1124 males with a BA to the 2614 males without a BA by taking all $2{,}938{,}136 = 1124 \times 2614$ differences. The boxplot after matching describes one number for each of the 829 matched sets, namely, the BA-minus-no-BA difference in means within a matched pair or triple. After matching, the differences are close to zero.

ERROR-FREE MILESTONES 15TABLE 1
*Estimates of percent returns to an additional year of education, using degree-scaled (DS) or self-reported (SRa) schooling. Least squares and m-estimation make no correction for measurement error. The two milestone methods use the BA as a milestone for 16 years of education, that is, as an instrumental variable. All methods adjust for four covariates. The table gives the point estimate, large sample 95% confidence interval and a standard error (se), which for the milestone estimate is the length of the 95% interval divided by $2 \times 1.96$*

| Method | Sample size | Variable | $\widehat{\beta}$ | se | 95% CI |
|---|---|---|---|---|---|
| Least squares | 3306 | DS | 0.035 | 0.0040 | $[0.027, 0.043]$ |
| Least squares | 3306 | SRa | 0.023 | 0.0036 | $[0.016, 0.030]$ |
| $m$-estimation | 3306 | DS | 0.038 | 0.0027 | $[0.032, 0.043]$ |
| $m$-estimation | 3306 | SRa | 0.030 | 0.0025 | $[0.026, 0.035]$ |
| Milestone Wilcoxon | 2248 | DS | 0.044 | 0.0036 | $[0.037, 0.051]$ |
| Milestone Wilcoxon | 2248 | SRa | 0.041 | 0.0036 | $[0.034, 0.048]$ |
| Milestone TSLS | 3306 | DS | 0.038 | 0.0045 | $[0.029, 0.047]$ |
| Milestone TSLS | 3306 | SRa | 0.035 | 0.0043 | $[0.027, 0.044]$ |

3.2. *Inference about economic returns to education.* Table 1 contrasts the eight estimates of economic returns to additional years of education, measured using log wages in 1974. Of the eight estimates, four are based on the better degree scaled education, DS, and four are based on self-report, SRa. Two methods, least squares and $m$-estimation (with R's defaults), make no correction for errors of measurement in SRa, whereas the third and fourth methods use the BA as a milestone for 16 years. In the third method, as described in Section 2.5, the special permutation distribution of Wilcoxon's signed rank statistic is used. The fourth method uses two-stage least squares with the milestone as the instrumental variable; however, conventional two-stage least squares actually requires more than (1) and (2), whereas these assumptions suffice for the Wilcoxon method. If DS were free of measurement error and SRa were prone to measurement error, then least squares and $m$-estimation applied to SRa would be inconsistent, but, assuming that a linear model for the covariates $\mathbf{x}_{ij}$ holds, the same methods applied to DS would be consistent. Table 1 asks the following: Which methods give similar answers with both DS and SRa?

Although the methods differ in several respects, not solely the use of the milestone, and although sampling variability creates some ambiguity, it does appear that (i) use of SRa in least squares or $m$-estimation yielded a lower estimated return to education, and (ii) using the milestone, DS and SRa produced similar results. Using the fallible self-report, SRa, the 95% confidence interval from $m$-estimation is $[0.026, 0.035]$, whereas using the milestone with the Wilcoxon procedure, it is $[0.034, 0.048]$, so these intervals barely overlap.

Although two-stage least squares used more observations than the Wilcoxon procedure, its confidence intervals were longer, perhaps because log (income) does not have a Gaussian distribution [see Imbens and Rosenbaum (2005), Figure 2(b)], or perhaps because of the remarkable property noted by Sen (1968) which is directly relevant to (2) when $F_{d_{ij}}$ varies with $d_{ij}$.

## 4. Multiple milestones.

4.1. *Definition and model: partition and reflection symmetry.* In this section we extend the model in Section 2 in two ways. First, we allow for multiple milestones for one variable, for instance, for years of education, twelve years for a high school diploma and sixteen years for a BA degree. In the WLS all respondents completed high school with a high school degree, so this milestone is not available. Second, we allow for several variables, each with at least one milestone. In the example in Section 5, using the WLS data, we will estimate the partial slopes for years of education and months of military service, using the BA as a milestone for sixteen years of education and no military service as a milestone for months of military service.

In contrast to Section 2, there is now a $P$-dimensional fixed vector $\mathbf{d}_{ij} = (d_{ij1}, \ldots, d_{ijP})$ of true but unobserved doses, a fallible, random $P$-dimensional observed dose $\mathbf{D}_{ij}$. In Section 5 $\mathbf{d}_{ij} =$ (years of education, months of military service). Write $\mathcal{S}_P$ for the set of $P$-dimensional, continuous multivariate distributions that are symmetric about $\mathbf{0}$, in the sense that if $\boldsymbol{\xi} \sim F \in \mathcal{S}_P$, then $\boldsymbol{\xi}$ and $-\boldsymbol{\xi}$ have the same distribution; see Sen and Puri (1967), Snijders (1981) or Neuhaus and Zhu (1998). The model is

$$
\begin{aligned}
Y_{ij} &= \lambda(\mathbf{x}_{ij}) + \boldsymbol{\beta}^T \mathbf{d}_{ij} + \varepsilon_{ij}, \qquad \varepsilon_{ij} \overset{\text{i.i.d.}}{\sim} G \in \mathcal{C}, \\
\mathbf{D}_{ij} &= \mathbf{d}_{ij} + \boldsymbol{\xi}_{ij}, \qquad \boldsymbol{\xi}_{ij} \sim F_{\mathbf{d}_{ij}} \in \mathcal{S}_P,
\end{aligned}
\tag{5}
$$

where the $\varepsilon_{ij}$ and $\boldsymbol{\xi}_{ij}$ are mutually independent. The $P$ coordinates of $\boldsymbol{\xi}_{ij}$ may be dependent; for instance, exaggerating years of education may be correlated with exaggerating months of military service. From (5), the distribution of measurement errors, $F_{\mathbf{d}_{ij}}$, varies with the true dose $\mathbf{d}_{ij}$, but any linear combination of the components of the true dose $\eta^T \mathbf{D}_{ij}$ is always symmetrically distributed about its center or median, namely, $\eta^T \mathbf{d}_{ij}$. Matching is assumed to exactly control $\mathbf{x}$, so that, as in Section 1.4, two individuals, $j$ and $k$, in the same matched set, $i$, have $\mathbf{x}_{ij} = \mathbf{x}_{ik}$.

Write $\mathcal{D}$ for the set of possible values of $\mathbf{d}_{ij}$. The generalization of an error-free milestone is a mutually exclusive and exhaustive partition $\mathcal{D} = \mathcal{D}_1 \cup \cdots \cup \mathcal{D}_L$, with $\mathcal{D}_\ell \cap \mathcal{D}_{\ell'} = \varnothing$ for $\ell \neq \ell'$ such that $\mathbf{d}_{ij} \in \mathcal{D}_\ell \Leftrightarrow \mathbf{D}_{ij} \in \mathcal{D}_\ell$. Because $\mathbf{d}_{ij}$ is fixed, $\mathbf{D}_{ij}$ is observed, and $\mathbf{D}_{ij} \in \mathcal{D}_\ell$ if and only if $\mathbf{d}_{ij} \in \mathcal{D}_\ell$; it follows that membership in a particular $\mathcal{D}_\ell$ is fixed and known, even though $\mathbf{d}_{ij}$ is not observed. The case of a single milestone had $\mathbb{R} = \mathcal{D} = \mathcal{D}_1 \cup \mathcal{D}_2$ with $\mathcal{D}_1 = \{d : d < \kappa\}$, $\mathcal{D}_2 = \{d : d \geq \kappa\}$. It is assumed that the partitioning cuts each of the $P$ coordinates at least once, so that the partition $\mathcal{D} = \mathcal{D}_1 \cup \cdots \cup \mathcal{D}_L$ includes at least $2^P$ quadrants formed by these $P$ cuts, which implies $L \geq 2^P$. In the example in Section 5, $\mathbf{d} =$ (years of education, months of military service), and the partition is $\mathcal{D} = \mathcal{D}_1 \cup \cdots \cup \mathcal{D}_4$, where $\mathcal{D}_1$ is "no BA, no military service" $\mathcal{D}_2$ is "no BA, some military service," $\mathcal{D}_3$ is "BA, no military service" and $\mathcal{D}_4$ is "BA, some military service." Also, in asymptotics, as $I \to \infty$, it is assumed that the fraction of observations in $\mathcal{D}_\ell$ tends to a positive constant, $\phi_\ell > 0$, for each $\ell$, where $1 = \phi_1 + \cdots + \phi_L$.

Consider testing the null hypothesis $H_0 : \boldsymbol{\beta} = \boldsymbol{\beta}_0$ using $Y_{ij} - \boldsymbol{\beta}_0^T \mathbf{D}_{ij}$, comparing matched individuals, $j$ and $k$, in the same matched set $i$, where

$$
\begin{aligned}
V_{ijk}^{(\boldsymbol{\beta}_0)} &= (Y_{ij} - \boldsymbol{\beta}_0^T \mathbf{D}_{ij}) - (Y_{ik} - \boldsymbol{\beta}_0^T \mathbf{D}_{ik}) \\
&= \boldsymbol{\beta}^T (\mathbf{d}_{ij} - \mathbf{d}_{ik}) - \boldsymbol{\beta}_0^T (\mathbf{D}_{ij} - \mathbf{D}_{ik}) + (\varepsilon_{ij} - \varepsilon_{ik}) \\
&= (\boldsymbol{\beta} - \boldsymbol{\beta}_0)^T (\mathbf{d}_{ij} - \mathbf{d}_{ik}) - \boldsymbol{\beta}_0^T (\xi_{ij} - \xi_{ik}) + (\varepsilon_{ij} - \varepsilon_{ik}),
\end{aligned}
\tag{6}
$$

which is symmetric about zero if $H_0 : \boldsymbol{\beta} = \boldsymbol{\beta}_0$ is true. If $H_0$ is false, then $V_{ijk}^{(\boldsymbol{\beta}_0)}$ is symmetric about $(\boldsymbol{\beta} - \boldsymbol{\beta}_0)^T (\mathbf{d}_{ij} - \mathbf{d}_{ik})$. Of course, (6) is the multivariate analogue of (4).

4.2. *Optimal nonbipartite matching; vector of signed-rank statistics.* We focus on the case of matched pairs, $n_i = 2$, selected to ensure that $\mathbf{x}_{i1}$ and $\mathbf{x}_{i2}$ are as close as possible and that if $\mathbf{d}_{i1} \in \mathcal{D}_\ell$, then $\mathbf{d}_{i2} \notin \mathcal{D}_\ell$. Define a distance, such as the Mahalanobis distance, between values of $\mathbf{x}$, and compute that distance for every possible pair of two individuals; however, if two individuals have $\mathbf{D}$ in the same $\mathcal{D}_\ell$, then replace that distance by $\infty$. With these distances, apply optimal nonbipartite matching to construct the pairs, as described in Lu and Rosenbaum (2004), thereby finding a pairing that minimizes the total distance within pairs on $\mathbf{x}$ subject to the constraint that paired individuals are in different $\mathcal{D}_\ell$. Algorithms for optimal nonbipartite matching are discussed by Galil (1986), Derigs (1988) and Cook and Rohe (1999).

Recall that $\mathbf{d}_{ij} \in \mathcal{D}_\ell \Leftrightarrow \mathbf{D}_{ij} \in \mathcal{D}_\ell$. If $\mathbf{D}_{i1} \in \mathcal{D}_\ell$ and $\mathbf{D}_{i2} \in \mathcal{D}_{\ell'}$, for $p = 1, \ldots, P$, define $z_{ip} = 1$ if $\mathbf{d} \in \mathcal{D}_\ell$, $\mathbf{d}' \in \mathcal{D}_{\ell'}$ implies $d_p > d'_p$, $z_{ip} = -1$ if $\mathbf{d} \in \mathcal{D}_\ell$, $\mathbf{d}' \in \mathcal{D}_{\ell'}$ implies $d_p < d'_p$, $z_{ip} = 0$ if $\mathbf{d} \in \mathcal{D}_\ell$, $\mathbf{d}' \in \mathcal{D}_{\ell'}$ does not itself determine the ordering of $d_p$ and $d'_p$. For instance, in Section 5, if $\mathbf{D}_{i1} \in \mathcal{D}_2 =$ "no BA, some military service," $\mathbf{D}_{i2} \in \mathcal{D}_4 =$ "BA, some military service," then $z_{i1} = -1$ and $z_{i2} = 0$. Write $\mathbf{z}_i = (z_{i1}, \ldots, z_{iP})^T$. Notice that the $\mathbf{z}_i$ are determined by the fixed events $\mathbf{d}_{ij} \in \mathcal{D}_\ell \Leftrightarrow \mathbf{D}_{ij} \in \mathcal{D}_\ell$, so the $\mathbf{z}_i$ are fixed.

Consider the hypothesis, $H_0 : \boldsymbol{\beta} = \boldsymbol{\beta}_0$, and let $r_{i, \boldsymbol{\beta}_0}$ be the rank of $|V_{i12}^{(\boldsymbol{\beta}_0)}|$ in (6), let $s_{i, \boldsymbol{\beta}_0} = \text{sign}(V_{i12}^{(\boldsymbol{\beta}_0)})$, and define the $P$-dimensional vector of signed

rank statistics,

$$\mathbf{T}_{\boldsymbol{\beta}_0} = \sum_{i=1}^{I} \mathbf{z}_i r_{i,\boldsymbol{\beta}_0} s_{i,\boldsymbol{\beta}_0}.$$

If $H_0: \boldsymbol{\beta} = \boldsymbol{\beta}_0$ were true, then $s_{i,\boldsymbol{\beta}_0} = \text{sign}(V_{i12}^{(\boldsymbol{\beta}_0)})$ and $|V_{i12}^{(\boldsymbol{\beta}_0)}|$ would be independent; again, see Wolfe (1974), Corollary 2.1. Consider the conditional distribution of $\mathbf{T}_{\boldsymbol{\beta}_0}$ given the $|V_{i12}^{(\boldsymbol{\beta}_0)}|$; under $H_0: \boldsymbol{\beta} = \boldsymbol{\beta}_0$, this distribution has $E(\mathbf{T}_{\boldsymbol{\beta}_0}) = \mathbf{0}$ and $P \times P$ covariance matrix

(7) $$\text{var}(\mathbf{T}_{\boldsymbol{\beta}_0}) = \sum_{i=1}^{I} r_{i,\boldsymbol{\beta}_0}^2 \mathbf{z}_i \mathbf{z}_i^T.$$

This variance formula (7) depends upon the continuous distribution of $V_{ijk}^{(\boldsymbol{\beta}_0)}$, which ensures $|V_{i12}^{(\boldsymbol{\beta}_0)}| > 0$ and $|s_{i,\boldsymbol{\beta}_0}| = 1$ with probability one. In the presence of ties, use average ranks for tied ranks, and use $\text{var}(\mathbf{T}_{\boldsymbol{\beta}_0}) = \sum_{i=1}^{I} |s_{i,\boldsymbol{\beta}_0}| r_{i,\boldsymbol{\beta}_0}^2 \mathbf{z}_i \mathbf{z}_i^T$. If $H_0: \boldsymbol{\beta} = \boldsymbol{\beta}_0$ were true, then $\mathbf{T}_{\boldsymbol{\beta}_0}^T \{\text{var}(\mathbf{T}_{\boldsymbol{\beta}_0})\}^{-1} \mathbf{T}_{\boldsymbol{\beta}_0}$ would tend to the chi-square distribution on $P$ degrees of freedom [Sen and Puri (1967)], and from this a confidence set for $\boldsymbol{\beta}$ is found by inverting the test. The point estimate of $\boldsymbol{\beta}$ minimizes $\mathbf{T}_{\boldsymbol{\beta}_0}^T \{\text{var}(\mathbf{T}_{\boldsymbol{\beta}_0})\}^{-1} \mathbf{T}_{\boldsymbol{\beta}_0}$ as a function of $\boldsymbol{\beta}_0$.

**5. An example with multiple milestones: returns to education and military service.** In the WLS data of Section 1.6, with $\mathbf{d}_{ij} = $ (years of education in DS, true months of military service), with the BA degree as a milestone for 16 years of education and with no military service as a milestone for zero years of service, the partial regression coefficients $\boldsymbol{\beta} = (\beta_{ed}, \beta_{ms})^T$ will be estimated from the fallible self report $\mathbf{D}_{ij} = $ (years of education in SRa, measured months of military service). As in Section 3, men were matched for a 4-dimensional $\mathbf{x}$ consisting of IQ in high school, father's education in years, mother's education in years and the population size of the town in which the individuals attended high school. From the 3738 men, we formed 1000 pairs of two men by optimal nonbipartite matching, as described in Section 4.2, where the distance was the Mahalanobis distance computed from the ranks of the four variables and from two indicators for missing parental education. The matching resulted in 230 pairs whose members differ on which side of both the BA and military service milestones they are on, 199 pairs whose members differ only on which side of the BA milestone they are on and 571 pairs whose members differ only on which side of the military service milestone they are on. Among the 230 pairs whose members differed on both BA and military service, in 143 pairs, one member had a BA and no military service and the other member had no BA and military service, and in 87 pairs, one member had a BA and military service and the other member had neither a BA nor military service. Boxplots similar to Figure 1 but not included show that the differences on **x** are nearly zero within matched pairs.

We now compare the least squares, $m$-estimate and milestone estimates of percent returns to an additional year of education and an additional month of military service. The least squares estimates and $m$-estimates regress log wages on self reported education (SRa), months of military service, IQ, father's education, mother's education and scaled hometown population and lose 445 men due to missing data on parent's education or months of military service, leaving 3293 men. The milestone estimates are based on the nonbipartite matching described above of 1000 pairs of two men, matching missing data to missing data, using the methods in Section 4.2.

Figure 4 plots the three 95% confidence sets $\boldsymbol{\beta} = (\beta_{ed}, \beta_{ms})^T$. As in Section 3, the milestone method suggests the returns to education, $\beta_{ed}$, are higher than the two regression methods that ignore measurement error. Specifically, for least squares $\widehat{\beta}_{ed}$ is about a 2% increase in earnings per year of education, for $m$-estimation $\widehat{\beta}_{ed}$ is about 3% per year, and for the milestone method $\widehat{\beta}_{ed}$ is about 4% per year. For military service, the milestone method suggests $\beta_{ms}$ might be zero, whereas the regression methods suggest reduced earnings. Table 2 presents numerical results. The confidence intervals for the milestone

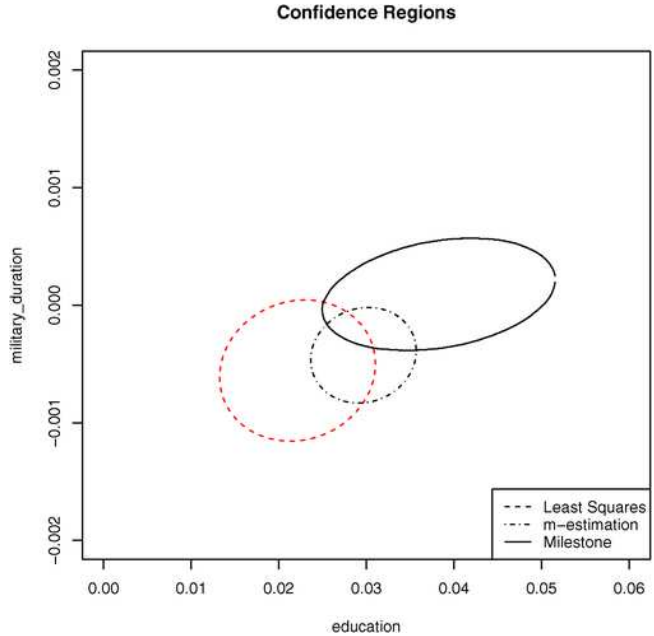

FIG. 4. *95% confidence sets for $(\beta_{ed}, \beta_{ms})^T$ by three methods.*

TABLE 2
*Estimates of the percent returns to an additional year of education (ed) or an additional month of military service (ms). The two stage least squares estimates use receipt of a BA and whether the man served in the military as instrumental variables*

| Method | Sample size | $\widehat{\beta}_{ed}$ | se | 95% CI |
|---|---|---|---|---|
| Education | | | | |
| Least squares | 3293 | 0.02214 | 0.00363 | [0.01502, 0.02927] |
| $m$-estimation | 3293 | 0.02967 | 0.00246 | [0.02485, 0.03449] |
| Milestone | 2000 | 0.03820 | 0.00679 | [0.02495, 0.05155] |
| TSLS | 3293 | 0.03587 | 0.00436 | [0.02733, 0.04441] |
| **Method** | **Sample size** | $\widehat{\beta}_{ms}$ | **se** | **95% CI** |
| Military | | | | |
| Least squares | 3293 | $-0.00056$ | 0.00025 | $[-0.00104, -0.00007]$ |
| $m$-estimation | 3293 | $-0.00042$ | 0.00017 | $[-0.00075, -0.00009]$ |
| Milestone | 2000 | 0.00009 | 0.00024 | $[-0.00039, 0.00057]$ |
| TSLS | 3293 | 0.00032 | 0.00046 | $[-0.00058, 0.00121]$ |

method are projections of the confidence set, so their simultaneous coverage is 95%.

**6. Summary.** For use with measurement error, there are several methods for using strong, valid instrumental variables [e.g., Cheng and Van Ness (1999), Section 4.2], but few methods for constructing them. Error-free milestones in error prone measurements create instrumental variables. In the Alzheimer's disease example in Section 1.2, the dose of anesthesia is measured with error, except for the zero doses of patients who did not have surgery. In the combat exposure scale example in Section 1.3, certain points on the scale are anchored by military records, while others depend on self report and memory, the latter being far more prone to error. In surveys in Section 1.4, a scaled response with aspects prone to error because of subtle operational definitions or memory lapses may sometimes be anchored by events that are difficult to misunderstand or forget, such as events marked by public ceremony or official sanction. Although various generalizations were mentioned, the discussion has focused on a predictor that has errors which are symmetric about zero yet respect a milestone, and in this case, exact, nonparametric inference was developed.

## REFERENCES

BRENNER, H. and GEFELLER, O. (1993). Use of the positive predictive value to correct for disease misclassification in epidemiologic studies. *Am. J. Epidemiol.* **138** 1007–1015.

CARROLL, R. J., RUPPERT, D. and STEFANSKI, L. A. (1995). *Measurement Error in Nonlinear Models*. Chapman & Hall, London. MR1630517

Cheng, C.-L. and Van Ness, J. W. (1999). *Statistical Regression With Measurement Error. Kendall's Library of Statistics* **6**. Oxford, New York.

Cochran, W. G. (1968). The effectiveness of adjustment by subclassification in removing bias in observational studies. *Biometrics* **24** 295–313. MR0228136

Cook, W. and Rohe, A. (1999). Computing minimum-weight perfect matchings. *INFORMS J. Comput.* **11** 138–148. C code is available at http://www2.isye.gatech.edu/~wcook/. MR1696029

Dee, T. S., Evans, W. N. and Murray, S. E. (1999). Data watch: Research data in the economics of education. *J. Econ. Perspect.* **13** 205–216.

Derigs, U. (1988). Solving non-bipartite matching problems via shortest path techniques. *Ann. Oper. Res.* **13** 225–261. MR0950993

Eckenhoff, R. G., Johansson, J. S., Wei, H., Carnini, A., Kang, B., Wei, W., Pidikiti, R., Keller, J. M. and Eckenhoff, M. F. (2004). Inhaled anesthetic enhancement of amyloid-$\beta$ oligomerization and cytotoxicity. *Anesthesiology* **101** 703–709.

Fuller, W. A. (1987). *Measurement Error Models*. Wiley, New York. MR0898653

Galil, Z. (1986). Efficient algorithms for finding maximum matching in graphs. *ACM Comp. Surv.* **18** 23–38. MR0859965

Gu, X. S. and Rosenbaum, P. R. (1993). Comparison of multivariate matching methods: Structures, distances and algorithms. *J. Comput. Graph. Statist.* **2** 405–420.

Hansen, B. B. (2004). Full matching in an observational study of coaching for the SAT. *J. Amer. Statist. Assoc.* **99** 609–618. MR2086387

Hansen, B. B. (2007). Optmatch: Flexible, optimal matching for observational studies. *R News* **7** 18–24.

Hansen, B. and Klopfer, S. O. (2006). Optimal full matching and related designs via network flows. *J. Comput. Graph. Statist.* **15** 609–627. MR2280151

Haviland, A., Nagin, D. and Rosenbaum, P. R. (2007). Combining propensity score matching and group-based trajectory analysis in an observational study. *Psychol. Methods* **12** 247–267.

Hettmansperger, T. P. and McKean, J. W. (1998). *Robust Nonparametric Statistical Methods*. Wiley, New York. MR1604954

Hodges, J. L. and Lehmann, E. L. (1963). Estimates of location based on ranks. *Ann. Math. Statist.* **34** 598–611. MR0152070

Huber, P. J. (1981). *Robust Statistics*. Wiley, New York. MR0606374

Imbens, G. and Rosenbaum, P. R. (2005). Robust, accurate confidence intervals with a weak instrument: Quarter of birth and education. *J. Roy. Statist. Soc. Ser. A* **168** 109–126. MR2113230

Johnson, T., Monk, T., Rasmussen, L. S. et al. (2002). Postoperative cognitive dysfunction in middle-aged patients. *Anesthesiology* **96** 1351–1357.

Kendall, M. G. and Stuart, A. (1973). *The Advanced Theory of Statistics*, *Volume 2: Inference and Relationship*, 3rd ed. Hafner, New York. MR0474561

Lu, B. and Rosenbaum, P. R. (2004). Optimal pair matching with two control groups. *J. Comput. Graph. Statist.* **13** 422–434. R code is available at http://cph.osu.edu/divisions/biostatistics/biostatsfacstaff/lub/. MR2063993

Lund, M., Foy, D., Sipprelle, C. and Strachan, A. (1984). The combat exposure scale: A systematic assessment of trauma in the Vietnam War. *J. Clin. Psychol.* **40** 1323–1328.

Lyberg, L. (1997). *Survey Measurement and Process Quality*. Wiley, New York.

Madansky, A. (1959). The fitting of straight lines when both variables are subject to error. *J. Amer. Statist. Assoc.* **54** 173–205. MR0102875

Neuhaus, G. and Zhu, L. X. (1998). Permutation tests for reflected symmetry. *J. Multivar. Anal.* **67** 129–153. MR1659215

Neyman, J. and Scott, E. L. (1951). On certain methods of estimating the linear structural relation. *Ann. Math. Statist.* **22** 352–361. MR0043416

Rosenbaum, P. R. (1991). A characterization of optimal designs for observational studies. *J. Roy. Statist. Soc. Ser. B* **53** 597–610. MR1125717

Rosenbaum, P. R. (2005). Exact, nonparametric inference when doses are measured with random errors. *J. Amer. Statist. Assoc.* **100** 511–518. MR2160555

Rosenbaum, P. R. (2006). Differential effects and generic biases in observational studies. *Biometrika* **93** 573–586. MR2261443

Sen, P. K. (1968). On a further robustness property of the test and estimator based on Wilcoxon's signed rank statistic. *Ann. Math. Statist.* **39** 282–285. MR0221697

Sen, P. K. and Puri, M. (1967). On rank order tests for location in the multivariate one sample problem. *Ann. Math. Statist.* **38** 1216–1228. MR0212954

Silber, J. H., Rosenbaum, P. R., Zhang, X. and Even-Shoshan, O. (2007). Estimating anesthesia and surgical time from medicare anesthesia claims. *Anesthesiology* **106** 346–355.

Singer, B., Ryff, C. D., Carr, D. and Magee, W. J. (1998). Linking life histories and mental health. *Sociol. Methodol.* **28** 1–51.

Snijders, T. (1981). Rank tests for bivariate symmetry. *Ann. Statist.* **9** 1087–1095. MR0628764

Sudman, S. and Bradburn, N. M. (1986). *Asking Questions.* Jossey-Bass, San Francisco.

Tourangeau, R., Rips, L. J. and Rasinski, K. (2000). *The Psychology of Survey Response.* Cambridge Univ. Press, New York.

Trotter, W. T. (1992). *Combinatorics and Partially Ordered Sets: Dimension Theory.* Johns Hopkins Univ. Press, Baltimore, MD. MR1169299

Wald, A. (1940). The fitting of straight lines if both variables are subject to error. *Ann. Math. Statist.* **11** 284–300. MR0002739

Warren, J. R., Sheridan, J. T. and Hauser, R. M. (2002). Occupational stratification across the life course: Evidence from the Wisconsin Longitudinal Study. *Am. Sociol. Rev.* **67** 432–455.

Wolfe, D. A. (1974). A characterization of population weighted symmetry and related results. *J. Amer. Statist. Assoc.* **69** 819–822. MR0426239

Department of Statistics
The Wharton School
University of Pennsylvania
473 Jon M. Huntsman Hall
3730 Walnut Street
Philadelphia, Pennsylvania 19104-6340
USA
E-mail: dsmall@wharton.upenn.edu
       rosenbaum@stat.wharton.upenn.edu